\documentclass[conference]{IEEEtran}
\IEEEoverridecommandlockouts
\usepackage{cite}
\usepackage{amsmath,amssymb,amsfonts}

\usepackage{graphicx}
\usepackage{textcomp}
\usepackage{xcolor}
\usepackage{comment}
\usepackage[ruled,vlined]{algorithm2e}%
\usepackage[flushleft]{threeparttable}%

\def\BibTeX{{\rm B\kern-.05em{\sc i\kern-.025em b}\kern-.08em
    T\kern-.1667em\lower.7ex\hbox{E}\kern-.125emX}}
\begin{document}

\title{QUADRO: A Hybrid Quantum Optimization Framework for Drone Delivery\\
\thanks{James B. Holliday acknowledges support from J.B. Hunt Transport Inc.}
}

\author{\IEEEauthorblockN{James B. Holliday\IEEEauthorrefmark{1}}
\IEEEauthorblockA{\textit{Dept. of EECS} \\
\textit{University of Arkansas}\\
\textit{J.B. Hunt Inc.}\\
Fayetteville, AR, USA \\
jbhollid@uark.edu}
\and
\IEEEauthorblockN{Darren Blount\IEEEauthorrefmark{1}, Hoang Quan Nguyen\IEEEauthorrefmark{1}}
\IEEEauthorblockA{\textit{Dept. of EECS} \\
\textit{University of Arkansas}\\
Fayetteville, AR, USA \\
\{darrenb, hn016\}@uark.edu}
\and
%\IEEEauthorblockN{Hoang Quan Nguyen\IEEEauthorrefmark{1}}
%\IEEEauthorblockA{\textit{CVIU Lab} \\
%\textit{University of Arkansas}\\
%Fayetteville, AR, USA \\
%hn016@uark.edu}
%\and
\IEEEauthorblockN{Samee U. Khan}
\IEEEauthorblockA{\textit{Dept. of ECE} \\
\textit{Mississippi State University}\\
Starkville, MS, USA \\
skhan@ece.msstate.edu}
\and
\IEEEauthorblockN{Khoa Luu\IEEEauthorrefmark{1}}
\IEEEauthorblockA{\textit{Dept. of EECS} \\
\textit{University of Arkansas}\\
Fayetteville, AR, USA \\
khoaluu@uark.edu}
\IEEEauthorblockA{
    \IEEEauthorrefmark{1}\tt\scriptsize{https://uark-cviu.github.io/}
    }
}

\maketitle

\begin{abstract}
Quantum computing holds transformative potential for optimizing large-scale drone fleet operations, yet its near-term limitations necessitate hybrid approaches blending classical and quantum techniques. This work introduces Quantum Unmanned Aerial Delivery Routing Optimization (QUADRO), a novel hybrid framework addressing the Energy-Constrained Capacitated Unmanned Aerial Vehicle Routing Problem and the Unmanned Aerial Vehicle Scheduling Problem. By formulating these challenges as Quadratic Unconstrained Binary Optimization problems, QUADRO leverages the Quantum Approximate Optimization Algorithm for routing and scheduling, enhanced by classical heuristics and post-processing. We minimize total transit time in routing, considering payload and battery constraints, and optimize makespan scheduling across various drone fleets. Evaluated on adapted Augerat benchmarks (16–51 nodes), QUADRO competes against classical and prior hybrid methods, achieving scalable solutions with fewer than one hundred qubits. The proposed results underscore the viability of hybrid quantum-classical strategies for real-world drone logistics, paving the way for quantum-enhanced optimization in the Noisy Intermediate Scale Quantum era.
\end{abstract}


\section{Introduction}
The transformative potential of drones, or unmanned aerial vehicles (UAVs), spans industries like delivery, emergency response, and surveillance, yet commanding large fleets poses complex combinatorial optimization challenges. Quantum computing (QC) is near the capability to address these applications at scales beyond classical computers, but in the near term, QC is yet to be completed. Meanwhile, hybrid approaches that combine classical and quantum strengths show promise. While standalone QC remains limited by qubit quantity and stability, hybrid algorithms break down real-world problems, such as drone routing with capacity and battery constraints, into manageable parts, solvable with fewer than one hundred qubits. This intersection of QC and drone optimization promises efficient fleet management, pushing beyond toy problems toward practical applications.

\begin{figure}[t]
    \centering
    \includegraphics[width=0.98\linewidth]{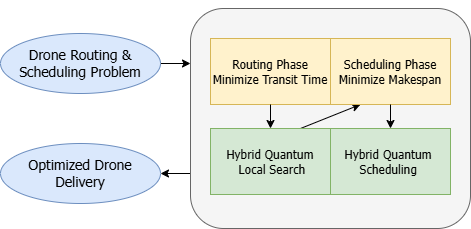}
    \caption{QUADRO: Framework for Drone Routing and Scheduling}
    \label{fig:teaser}
\end{figure}

\begin{figure}[t]
    \centering
    \includegraphics[width=0.98\linewidth]{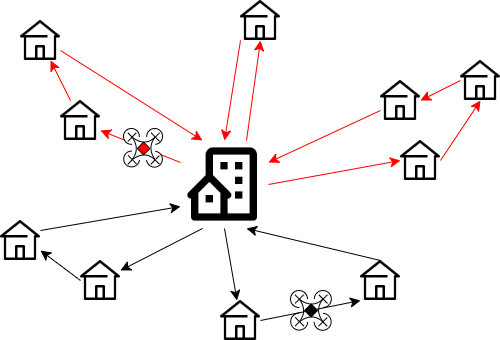}
    \caption{Drone delivery example with a two-drone fleet. The corresponding colored drone services colored routes. The central location is the depot, where the drones are loaded and recharged between routes.}
    \label{fig:timewindowroute}
\end{figure}

\textbf{Our Contributions in this Work:} 
We introduce a new Quantum Unmanned Aerial Delivery Routing Optimization (QUADRO), a hybrid quantum-classical framework, to solve the drone routing and scheduling problem, refer to Fig. \ref{fig:timewindowroute}. We adapt the Vehicle Routing Problem (VRP) to the drone routing problem by considering payload and battery capacities. Instead of minimizing the distance of the formulated routes, we minimize route transit time. We develop a multi-start hybrid routing algorithm with a quantum local search component to create the routes. Once the optimized routes are generated, we optimally schedule them on various sized fleets of drones to minimize the overall time needed to make the deliveries. Again, we utilize a quantum algorithm to schedule the routes to the fleet of drones optimally. We model the unique problems for both quantum algorithms as quadratic unconstrained binary optimization problems (QUBO) and solve them using QC. Fig. \ref{fig:teaser} visualizes our framework. We demonstrate the capability of our framework on different problem scales from sixteen deliveries to one hundred deliveries using fleets of two, three, and four drones. We compare our results to other research and show that our framework is comparable to cutting-edge classical approaches. 

The remainder of this article is organized as follows. Section \ref{background} will review the background related to this research. Section \ref{formulations} shows problem definitions. Section \ref{method} introduces our algorithm. Section \ref{experiments} contains our experimental setup and results. Lastly, Section \ref{conclusion} is for our findings.

\section{Background}\label{background}

Optimization related to drone delivery has a growing research focus. The range of optimization challenges in the drone delivery space is sizeable. From route planning to energy management and payload optimization to fleet management \cite{sajid2022routing}, regulatory compliance, and weather and environmental factors \cite{gao2021weather, tran2022management}. Drone delivery can be assisted by vehicles \cite{boysen2018drone} and tall buildings in the delivery zone \cite{kim2020drone}. Drone scheduling \cite{torabbeigi2020drone} and drone path-planning \cite{shivgan2020energy} are also important problems.

Quantum computing offers a potential option, having demonstrated success with NP-hard problems like the Traveling Salesman Problem \cite{hoffman2013traveling} and Vehicle Routing Problem \cite{toth2002vehicle}. However, in the Noisy Intermediate-Scale Quantum (NISQ) era, QC’s scale remains limited by qubit stability and error correction \cite{preskill2018quantum}, restricting solutions to small instances.

Hybrid QC approaches, integrating classical and quantum techniques, have emerged as the most viable path to tackle more significant, realistic problems with QC \cite{feld2019hybrid, borowski2020new, sinno2023performance, holliday2024tabu, XBacWK2, XBacWK1, XBacjournal1, AdityaQIP}. These methods decompose complex optimization tasks into subproblems suited for near-term QC, reducing qubit demands significantly. Recent research highlights promising intersections with QC and drone optimization \cite{kumar2022futuristic}. Such advancements position hybrid QC as a viable approach for drone optimization, bridging the gap between theoretical potential and practical utility in the NISQ era.

\subsection{Drone Routing and Scheduling} \label{sec:problems}

Drone Routing and Scheduling extends traditional vehicle routing paradigms by incorporating the unique operational characteristics of drones.
For our research, we focused on a subset of possible challenges in this optimization field. Those are route planning, energy management, and fleet management. Putting these together into a single problem, we will attempt to optimize freight delivery to customers using a discrete fleet of drones to minimize the total time utilized. We accomplish this goal by creating routes assigned to drones and then scheduled to be delivered in sequence by the fleet. In our case, we will use a singular depot where the drones begin and end each route. When a drone finishes a route, it must recharge before starting the following route. In addition, all routes must be feasible for a single drone without recharging mid-route. The capacity of the drone cannot be exceeded in either payload weight or battery consumption. 

While the payload capacity constraint in vehicle routing is common, the energy consumption consideration is unique. Drones are powered by a battery that has a limited capacity. As the drone moves along a route, it depletes the battery, so any actions the drone needs to perform must not consume more energy than the battery contains. The weight of the drone will also impact battery consumption. The heavier the drone, the more energy it consumes while performing actions. In addition, the weight of the drone will impact transit time by decreasing velocity, so minimizing the weight along the route becomes a critical factor. Removing heavier payload items earlier in the route will lead to shorter transit times. Shorter times will improve the overall objective of minimizing the total time needed to meet the demands of all customers. This routing problem variant is the Energy-Constrained Capacitated Unmanned Aerial Vehicle Routing Problem (EC-CUAVRP).  

In addition to the challenges faced in routing, we add the fleet management challenge of route scheduling optimization. After the routing is complete, the routes must be operated by the drones. If the routes are scheduled poorly, the total time to make all the deliveries will be slower than the optimized schedule. The fleet of drones is limited in size, so in many cases, a single drone must operate multiple routes. Before a drone can start another route, it must recharge its battery. For our case, we will consider a homogenous fleet of drones, meaning all their capacities and characteristics are the same. This scheduling problem is called the Unmanned Aerial Vehicle Scheduling Problem (UAVSP).

\subsection{Quantum Approximate Optimization Algorithm} \label{sec:QC}

Quantum Approximate Optimization Algorithm (QAOA) is a quantum algorithm designed to solve combinational optimization problems \cite{farhi2014quantum}.
QAOA is inspired by the adiabatic quantum computing model, where a quantum system evolves gradually from an initial state to a ground state, presenting the solution to the optimization problem.
The algorithm consists of two leading operators, i.e., cost Hamiltonian $H_C$ encoding the objective function to be optimized and mixing Hamiltonian $H_B$, ensuring exploration. 

Given an optimization problem, the objective function is encoded as a cost Hamiltonian $H_C$, with the goal to minimize or maximize its eigenvalues.
The initial state is the equal superposition of all possible basis states as in Eqn. \eqref{eq:eq1}.
\begin{equation} \label{eq:eq1}
    |\psi_0\rangle = \frac{1}{\sqrt{2^n}} \sum_{x=0}^{n-1} |x\rangle,
\end{equation}
where $n$ is the number of qubits.
Then, the QAOA applies a sequence of alternating time evolution operators parameterized by time $\gamma_j$ and $\beta_j$ as in Eqn. \eqref{eq:eq2}.
\begin{equation} \label{eq:eq2}
\begin{split}
    U_C(\gamma_j) &= e^{-i \gamma_j H_C}, \\
    U_B(\beta_j) &= e^{-i \beta_j H_B},
\end{split}
\end{equation}
where $i$ is the imaginary part.
The operators are alternatively repeated $p$ times:
\begin{equation}
    |\psi_p(\gamma, \beta)\rangle = 
    \left( 
    \prod_{j=1}^p U_B(\beta_j) U_C(\gamma_j)
    \right) |\psi_0\rangle.
\end{equation}
After applying the quantum circuits, the quantum state is measured computationally, and the resulting bitstring corresponds to a possible solution to the optimization problem.
% To find the optimal solution, a classical optimizer is used to update $\gamma_j$ and $\beta_j$ until an approximately optimal solution is found.
A classical optimizer, e.g., gradient descent or Bayesian optimization, is applied to update $\gamma_j$ and $\beta_j$ until an approximately optimal solution is found.

By adjusting the parameters of these operations, QAOA seeks to find approximate solutions to optimization problems, including Max-Cut \cite{wang2018quantum}, traveling salesman \cite{qian2023comparative,ramezani2024reducing}, and scheduling problems \cite{choi2020quantum,kurowski2023application}.
Since exact solutions are often computationally expensive, QAOA provides efficient approximations due to the advances in quantum computing.

\subsection{Quadratic Unconstrained Binary Optimization (QUBO) to Quantum Approximate Optimization Algorithm (QAOA)}

In general, the QUBO problem can be formulated as:
\begin{equation}
    \min_{X} \sum_{j,k} Q_{j,k} X_j X_k + \sum_{j} b_{j} X_j + C
\end{equation}
where $X_j \in \{0, 1\}$.
We then replace the binary variables $X_j$ with a new set of variables $Z_j \in \{-1, 1\}$ through $X_j = \frac{1 - Z_j}{2}$.
By substituting $Z_j$ for $X_j$, we obtain a new equivalent formulation as in Eqn. \eqref{eq:eq5}.
\begin{equation} \label{eq:eq5}
    \min_{X} \sum_{j,k} Q^\prime_{j,k} Z_j Z_k + \sum_{j} b^\prime_{j} Z_j + C^\prime
\end{equation}
where
\begin{equation}
\begin{split}
    Q^\prime_{j,k} &= \frac{Q_{j,k}}{4} \\
    b^\prime_j &= - \frac{b_j}{2} - \frac{1}{4} \sum_k (Q_{j,k} + Q_{k,j}) \\
    C^\prime &= C + \frac{1}{4} \sum_{j,k} Q_{j,k} + \frac{1}{2} \sum_{j} b_j
\end{split}
\end{equation}
To obtain the quantum formulation of the problem, we promote $Z_j$ variables to a Pauli-Z matrix $\sigma_j^Z$ defined as in Eqn. \eqref{eq:eq6}.
\begin{equation} \label{eq:eq6}
    \sigma_j^Z = \begin{pmatrix}
        1 & 0 \\
        0 & -1
    \end{pmatrix}
\end{equation}
Then we can replace $Z_j$ with $\sigma_j^Z$ to obtain the following Hamiltonian:
\begin{equation}
    H_C = \sum_{j,k} Q^\prime_{j,k} \sigma_j^Z \sigma_k^Z + \sum_{j} b^\prime_{j} \sigma_j^Z
\end{equation}
Note that since $C^\prime$ is a constant, we can remove this term.

\subsection{Related Work}
The intersection of QC and drone optimization is an emerging field with a limited yet expanding body of research, focusing primarily on hybrid quantum-classical approaches to routing and scheduling challenges. This section reviews four pertinent studies that inform our proposed framework. Initially, a classical approach \cite{sajid2022routing} inspired our methodology, presenting a joint-optimization framework for UAV-assisted delivery. They employ a hybrid genetic-simulated annealing (HGSA) algorithm for routing and a UAV-oriented MinMin (UO-MinMin) algorithm for scheduling, combining genetic exploration with annealing exploitation and resource-aware scheduling, respectively. Tested on modified CVRP benchmarks, HGSA outperforms other classical approaches in travel time, energy efficiency, and scheduling performance. In contrast, \cite{vista2023hybrid} proposes a hybrid QC method for scheduling in a UAV-enabled IoT network, formulating a resource allocation problem as a QUBO solved via quantum annealing, augmented by classical processing to mitigate qubit constraints. \cite{hua2024quantum} addresses a drone-specific TSP variant allowing vertex revisits and asymmetric costs, integrating classical pre-processing with QUBO formulations solved by quantum and digital annealers, outperforming Gurobi in TSPLIB and simulated urban tests. Similarly, \cite{osaba2025solving} introduces a hybrid QC approach combining QAOA-based clustering and quantum annealing for routing, validated across three use cases (12–22 nodes) against Google OR-Tools. These studies collectively address routing, scheduling, and real-world constraints—considerations that the QUADRO framework unifies into a single, integrated solution for fleet-based drone delivery, distinguishing it by optimizing both aspects.

\section{Problem Formulation} \label{formulations}

The problem addressed in this work encompasses both drone routing and scheduling for efficient delivery operations, modeled over an Euclidean graph \( G = (O^+, C) \). Here, \( O^+ = \{o_0\} \cup O \) represents a set of \( n+1 \) nodes, where \( o_0 \) is the depot with no demand (\( r_0 = 0 \)), and \( O = \{o_i : 1 \leq i \leq n\} \) denotes \( n \) customer nodes, each with geometric coordinates \( (x_i, y_i) \) and positive demand \( r_i > 0 \). The edge set \( C = \{(o_i, o_j) : o_i, o_j \in O, i \neq j\} \) connects nodes with Euclidean distances \( d_{ij} \). A fleet of \( m \) homogeneous drones, \( U = \{u_k : 1 \leq k \leq m\} \), is tasked with fulfilling customer demands, where each drone has uniform weight \( W \), payload capacity \( L \), and battery capacity \( B \). The customer set is partitioned into \( M \) aerial routes forming a batch-of-routes (BoR) \( J = \{A_z : 1 \leq z \leq M\} \), with each route \( A_z = (o_0, o_j, o_l, \ldots, o_0) \) starting and ending at the depot. 

The drone routing and scheduling problem is governed by assumptions and constraints to ensure operational feasibility and efficiency. It is assumed that a single drone can serve multiple customers per aerial route, launching from a depot where its battery is fully recharged prior to departure, and customers are pre-assigned to drones before take-off. How the deliveries are performed is outside the consideration of this work. Constraints mandate that the total demand of customers assigned to a drone must not exceed its payload capacity \(L\), and the power consumed along a route must remain within the battery capacity \(B\). Additionally, each customer’s demand must be fulfilled exactly once by a single drone, ensuring every customer is visited precisely once, and all drone routes must originate and terminate at the depot, enforcing a closed-loop structure for each mission.

\subsection{Drone Routing Problem}

The drone routing problem focuses on determining optimal aerial routes to minimize total travel time across all drones. Each drone \( u_k \) carries a payload \( L^k_i \) at node \( o_i \), defined as the sum of demands of subsequent nodes in its assigned route \( A_z \):
\begin{equation}
L^k_i = \sum_{o_j \in A^k_{zi}} r_j \quad \forall o_i \in O^+, \forall A_z \in J, \forall u_k \in U
\label{eq:payload}
\end{equation}
where \( A^k_{zi} = (o_j, o_l, \ldots, o_0) \) is the remaining path from \( o_i \) in route \( A_z \). The decision variable \( X^k_{ij} \in \{0, 1\} \) indicates whether drone \( u_k \) travels from \( o_i \) to \( o_j \) (1 if true, 0 otherwise). Flight time between nodes is derived from velocity:
\begin{equation}
 v^k_{ij} = \frac{370 \phi \gamma (P_H - p_e)}{W + L^k_i} 
\end{equation}
where \(\gamma\) is the lift-to-drag ratio, \(\phi\) is the conversion efficiency of the motor and propeller, and \(p_e\) is the power consumption of on-board electronics. Assuming maximum power is utilized at all times during the flight \( p^k_{ij} = P_H \), and yielding:
\begin{equation}
t^k_{ij} = \frac{d_{ij} (W + L^k_i)}{370 \phi \gamma (P_H - p_e)} \quad \forall o_i, o_j \in O^+, o_i \neq o_j, \forall u_k \in U
\label{eq:flight_time}
\end{equation}
Energy consumption per edge is:
\begin{multline}
e^k_{ij} = P_H t^k_{ij} = \frac{P_H d_{ij} (W + L^k_i)}{370 \phi \gamma (P_H - p_e)} \\ \forall o_i, o_j \in O^+, o_i \neq o_j, \forall u_k \in U
\label{eq:energy_per_edge}
\end{multline}
Total energy for route \( A_z \) is:
\begin{equation}
E^k_z = \sum_{(o_i, o_j) \in A_z} e^k_{ij} X^k_{ij} + \xi \quad \forall A_z \in J, \forall u_k \in U
\label{eq:route_energy}
\end{equation}
where \(\xi\) is the energy needed for take-off, landing, and delivery. The objective minimizes total travel time, incorporating flight and incidental times (\( \tau \) for loading, take-off, landing, and delivery) can be formulated as Eqn. \eqref{eq:objective_routing}.
\begin{equation}
\text{Minimize } T = \sum_{A_z \in J} \left( \sum_{(o_i, o_j) \in A_z} t^k_{ij} X^k_{ij} + (M + 1) \tau \right)
\label{eq:objective_routing}
\end{equation}
subject to the following constraints:
\begin{equation}
\sum_{u_k \in U} \sum_{o_i, o_j \in O^+} X^k_{ij} = 1
\label{eq:visit_once}
\end{equation}
\begin{equation}
\sum_{o_j \in A^k_{zi}} r_j \leq L \quad \forall o_i \in O^+, \forall A_z \in J, \forall u_k \in U
\label{eq:payload_constraint}
\end{equation}
\begin{equation}
\sum_{(o_i, o_j) \in A_z} p^k_{ij} X^k_{ij} \leq B \quad \forall A_z \in J, \forall u_k \in U
\label{eq:battery_constraint}
\end{equation}
\begin{equation}
 \sum_{o_j \in O} X^k_{0j} = 1 \quad \forall u_k \in U
\label{eq:depot_departure}
\end{equation}
\begin{equation}
 \sum_{o_i \in O^+} X^k_{ih} - \sum_{o_j \in O^+} X^k_{hj} = 0 \quad \forall o_h \in O^+, \forall u_k \in U
\label{eq:customer_continuity}
\end{equation}
\begin{equation}
 \sum_{o_i \in O} X^k_{i0} = 1 \quad \forall u_k \in U
\label{eq:depot_return}
\end{equation}
\begin{equation}
  X^k_{ij} \in \{0, 1\} \quad \forall o_i, o_j \in O^+, \forall u_k \in U
\label{eq:route_binary}
\end{equation}

Constraint (\ref{eq:visit_once}) ensures that a drone visits each customer exactly once. Constraint (\ref{eq:payload_constraint}), the capacity constraint, guarantees that the total demand of customers on a route does not exceed the drone’s payload capacity. Likewise, constraint (\ref{eq:battery_constraint}) limits a drone’s power consumption on a route to stay within its battery capacity. Constraints (\ref{eq:depot_departure}), (\ref{eq:customer_continuity}), and (\ref{eq:depot_return}) are flow constraints that enforce departure from the depot, continuity through customers, and return to the depot, respectively. Constraint (\ref{eq:route_binary}) specifies the primary binary decision variable.

\subsection{Drone Scheduling Problem}
The drone scheduling problem extends this by assigning \( M \) aerial routes to \( m \) drones to minimize the maximum completion time across all routes. Each route \( A_z \) is atomic and executed non-preemptively by one drone, with decision variable \( Y^k_z \in \{0, 1\} \) indicating assignment of \( A_z \) to \( u_k \). Take-off/Delivery/Landing execution time \(T_z\) is uniform across all routes and drones. The drone ready time \( URT_k \) depends on prior schedules as in Eqn. \eqref{eq:uav_ready_time}.
\begin{equation}
URT_k = 
\begin{cases} 
0 & \text{if } Q_k \text{ is empty} \\
CT_w + \tau_c & \text{otherwise}
\end{cases}
\label{eq:uav_ready_time}
\end{equation}
where \( CT_w \) is the completion time of the last scheduled route in queue \( Q_k \), and \( \tau_c \) is the battery recharge time. Route ready time is defined as in Eqn. \eqref{eq:route_ready_time}.
\begin{equation}
RT_z = \min \{ URT_k : \forall u_k \in U \}
\label{eq:route_ready_time}
\end{equation}
Completion time for \( A_z \) is as Eqn. \eqref{eq:completion_time}.
\begin{equation}
CT_z = RT_z + T_z
\label{eq:completion_time}
\end{equation}
The makespan is as Eqn. \eqref{eq:makespan}.
\begin{equation}
MK^U_J = \max \{ CT_z, \quad \forall A_z \in J \}
\label{eq:makespan}
\end{equation}
The objective is to minimize the makespan as in Eqn. \eqref{eq:objective_scheduling}.
\begin{equation}
\text{Minimize } MK^U_J = \max \{ CT_z, \quad \forall A_z \in J \}
\label{eq:objective_scheduling}
\end{equation}
subject to the following constraints:
\begin{equation}
\sum_{u_k \in U} Y^k_z = 1 \quad \forall A_z \in J
\label{eq:route_assignment}
\end{equation}
\begin{equation}
\sum_{A_z \in J} \sum_{u_k \in U} Y^k_z = M
\label{eq:total_routes}
\end{equation}
\begin{equation}
  Y^k_z \in {0,1} \quad \forall A_z \in J, \forall u_k \in U
\label{eq:schedule_binary}
\end{equation}

The first constraint (\ref{eq:route_assignment}) ensures that each route is assigned to precisely one drone. The second constraint (\ref{eq:total_routes}) guarantees that all \(M\) routes are scheduled across the drones. The third constraint (\ref{eq:schedule_binary}) confirms that the decision variable is binary. 

Table~\ref{tab:uav_parameters} presents the parameter values employed in the equations above, sourced from a classical approach to the same problem.  Adapted from~\cite{sajid2022routing}, the parameters and formulations integrate routing and scheduling to optimize delivery efficiency under stringent resource and operational constraints, establishing a solid foundation for our hybrid methodology. By unifying these dual objectives—minimizing transit time and makespan --— this approach tackles the intricacies of drone logistics, facilitating the application of QC to enhance real-world drone delivery systems.

\begin{table}[t]
    \centering
    \caption{Drone Parameters Used in the Routing and Scheduling Framework}
    \label{tab:uav_parameters}
    \begin{tabular}{l c r}
        \hline
        \textbf{Parameter Description} & \textbf{Symbol} & \textbf{Value} \\
        \hline
        UAV Weight & \( W \) & 7.5 kg \\
        Payload Capacity of UAV & \( L \) & 2.5 kg \\
        Conversion Efficiency of Motor and Propeller & \( \phi \) & 0.5 \\
        Lift-to-Drag Ratio & \( \gamma \) & 3 \\
        Power Consumption of Electronics & \( p_e \) & 0.1 kWh \\
        Battery Capacity & \( B \) & 1.7 kW \\
        Maximum Rate of Power & \( P_H \) & 0.6 kW \\
        Recharging Time & \( \tau_c \) & 1.25 h \\
        Take-off/Landing/Loading/Dropping Packages & \( \tau \) & 0.15 h \\
        Take-off/Landing/Dropping Packages & \( \xi \) & 0.015 kWh \\
        \hline
    \end{tabular}
\end{table}

\begin{figure}[tb]
    \centering
    \includegraphics[width=0.98\linewidth]{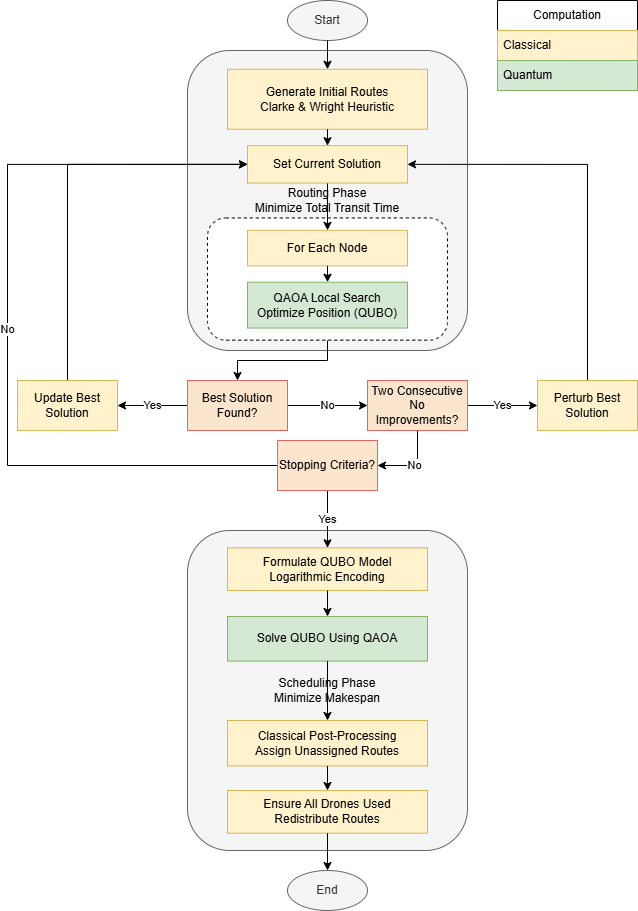}
    \caption{The proposed flow of our framework}
    \label{fig:flow}
\end{figure}

\section{Our Proposed Method} \label{method}

We propose the new Quantum Unmanned Aerial Delivery Routing Optimization (QUADRO), a hybrid quantum-classical framework, to solve the Energy-Constrained Capacitated Unmanned Aerial Vehicle Routing Problem (EC-CUAVRP) and Unmanned Aerial Vehicle Scheduling Problem (UAVSP), as defined in Section~\ref{formulations}, with the primary goal of maximizing QC utilization, visualized in Fig. \ref{fig:flow}. Drawing from our prior work~\cite{holliday2024tabu}, we leverage QC’s optimization potential by formulating small, simple QUBO problems tailored for near-term quantum hardware to tackle the combinatorial complexity of drone logistics. Our approach decomposes the problem into routing and scheduling phases, optimizing total transit time and makespan.  We use a multi-start strategy enhanced by a novel QAOA-based local search for routing. To ensure feasibility under payload and battery constraints, we offload constraint handling to classical components, reserving QC to explore routing solutions efficiently. We extend the scheduling component of QUADRO by formulating drone-route assignments as a QUBO problem and evaluating two distinct solution strategies: a purely quantum formulation and a hybrid quantum-classical approach. To address the scalability limitations of QAOA on large scheduling instances, we propose a logarithmic-encoded QUBO model for drone assignment. Our experiments across benchmark datasets demonstrate that this hybrid method maintains solution feasibility while scaling to larger problem sizes, supporting QUADRO’s goal of maximizing quantum resource utility through modular decomposition and targeted QUBO formulations.

\subsection{Routing Phase}

In the prior work~\cite{holliday2024tabu}, the Hybrid Quantum Tabu Search (HQTS) was developed for routing problems, combining a classical tabu search component with a QUBO formulation solved via quantum annealing. For this study, we shifted away from HQTS to devise a new hybrid routing method, driven by the goal of increasing quantum computing utilization and broadening the diversity of hybrid approaches in the research landscape. This routing phase aims to construct \( M \) feasible aerial routes that minimize the total transit time \( T \) (Eq.~\ref{eq:objective_routing}). The process begins with a multi-start heuristic, initializing a solution using Clarke and Wright’s savings algorithm~\cite{clarke1964scheduling}. A novel QUBO-based local search is applied from this starting point to refine the solution. For each node \( o \) in the set \( O^+ \), a QUBO is formulated to optimize the node’s position across all routes, minimizing the total transit time \( T \). Quantum computing, via the QAOA, explores alternative positions for \( o \), and the move yielding the lowest cost is selected to update the solution for the next iteration. This iterative process continues until two consecutive iterations yield no improvement, at this point, the best-known solution is perturbed, and the algorithm restarts, embodying the multi-start strategy.

The optimization targets a Batch of Routes \( J = \{A_z : 1 \leq z \leq M\} \), where each route \( A_z = (o_0, o_{z1}, o_{z2}, \ldots, o_{zn_z}, o_0) \) originates and terminates at the depot \( o_0 \), with customer nodes \( o_{zi} \) fixed except for one selected node, denoted \( o^* \), from a route \( A_s \in J \). The objective is to reassign \( o^* \) to its optimal position across all routes in \( J \), either within \( A_s \) at a different spot or in another route \( A_z, z \neq s \), to minimize \( T \) (Eq.~\ref{eq:objective_routing}), while payload and battery constraints are enforced classically before and after optimization. Binary variables \( X_{z,i} \in \{0, 1\} \) are defined, where \( X_{z,i} = 1 \) indicates that \( o^* \) is inserted into route \( A_z \) immediately before position \( i \) (between nodes \( o_{z,i-1} \) and \( o_{z,i} \), with \( o_{z,0} = o_0 \) and \( o_{z,n_z+1} = o_0 \)), and 0 otherwise; here, \( i = 1, 2, \ldots, n_z + 1 \) spans the possible insertion points in \( A_z \), with \( n_z \) as the number of customers in that route. To ensure \( o^* \) is placed exactly once, the following constraint is imposed:
\begin{equation}
    \sum_{z=1}^M \sum_{i=1}^{n_z + 1} X_{z,i} = 1
    \label{eq:qubo_single_assignment}
\end{equation}
For each route \( A_z \), inserting \( o^* \) at position \( i \) alters the transit time by replacing the edge \( (o_{z,i-1}, o_{z,i}) \) with \( (o_{z,i-1}, o^*) \) and \( (o^*, o_{z,i}) \). The flight time \( t^k_{ij} \) (Eq.~\ref{eq:flight_time}) depends on the payload \( L^k_i \) (Eq.~\ref{eq:payload}), which adjusts with \( o^* \)’s insertion due to its demand \( r^* \). Specifically, the time from \( o_{z,i-1} \) to \( o^* \) accounts for the distance and payload including \( r^* \), while the leg from \( o^* \) to \( o_{z,i} \) reflects a reduced payload post-delivery. These times are defined as follows:
\begin{equation}
    t_{z,i-1,*} = \frac{d_{z,i-1,*} (W + L^k_{z,i-1})}{370 \phi \gamma (P_H - p_e)}
    \label{eq:time_to_star}
\end{equation}
\begin{equation}
    t_{*,z,i} = \frac{d_{*,z,i} (W + L^k_{z,i-1} - r^*)}{370 \phi \gamma (P_H - p_e)}
    \label{eq:time_from_star}
\end{equation}
In these expressions, \( d_{z,i-1,*} \) is the distance from \( o_{z,i-1} \) to \( o^* \), \( L^k_{z,i-1} \) includes \( r^* \), and \( d_{*,z,i} \) is the distance from \( o^* \) to \( o_{z,i} \). The transit time change for \( A_z \) is then given by:
\begin{equation}
    \Delta T_{z,i} = t_{z,i-1,*} + t_{*,z,i} - t_{z,i-1,z,i}
    \label{eq:delta_transit}
\end{equation}
where \( t_{z,i-1,z,i} \) is the original edge time. The updated total transit time is expressed as:
\begin{equation}
    T' = T_0 - \sum_{(o_i, o_j) \in A_s} t^k_{ij} X^k_{ij} + \sum_{z=1}^M \sum_{i=1}^{n_z + 1} \Delta T_{z,i} X_{z,i}
    \label{eq:total_transit}
\end{equation}
with \( T_0 \) as the initial total time and the second term subtracting \( A_s \)’s contribution due to \( o^* \)’s removal. The QUBO Hamiltonian is formulated to minimize \( T' \) while enforcing the single-assignment constraint, as follows:
\begin{equation}
    H = \sum_{z=1}^M \sum_{i=1}^{n_z + 1} \Delta T_{z,i} X_{z,i} + \lambda \left( \sum_{z=1}^M \sum_{i=1}^{n_z + 1} X_{z,i} - 1 \right)^2
    \label{eq:qubo_hamiltonian}
\end{equation}
The first term sums the transit time changes weighted by \( X_{z,i} \), and the second term, with a penalty coefficient \( \lambda > 0 \) (chosen such that \( \lambda \gg \max |\Delta T_{z,i}| \)), ensures \( o^* \) is assigned exactly once. To construct this Hamiltonian, \( \Delta T_{z,i} \) is pre-calculated for all feasible \( z \), where feasibility excludes moves exceeding payload capacity. 

Payload constraints (Eqn.~\ref{eq:payload_constraint}) and battery constraints (Eqn.~\ref{eq:battery_constraint}) are managed classically. During QUBO formulation, potential moves violating payload limits are excluded, while battery feasibility is verified post-optimization, rejecting infeasible solutions and selecting the next-best option if needed. In practice, no instances arose where all QUBO solutions violated battery constraints, removing the need for additional classical heuristics.

\subsection{Scheduling Phase}

In alignment with the decomposition strategy of QUADRO, the scheduling phase focuses on assigning a set of predefined routes to a fleet of drones to minimize the overall makespan. This stage follows the route generation process and leverages quantum and classical methods to optimize routes distributed across available drones.
Specifically, each route must be assigned to precisely one drone, and our objective is to minimize the maximum load among all drones while satisfying operational constraints.

We introduce two QUBO-based formulations: a \textit{pure quantum} model and a \textit{hybrid quantum-classical} model, inspired by \cite{lucas2014ising}. The pure quantum formulation encodes all route-to-drone assignments using binary variables $x_{i,\alpha}$ and enforces assignment constraints and load balancing directly within the QUBO structure. This method offers a conceptually straightforward optimization path but quickly becomes intractable as the number of drones increases due to the quadratic explosion of binary variables. To overcome this limitation, we introduce a hybrid formulation that uses logarithmic encoding to reduce variable count and integrates classical post-processing to guarantee feasibility. 

In large-scale problems, the dimensionality of decision variables escalates rapidly, rendering exact methods computationally prohibitive. Recent advancements in quantum computing have spurred efforts to encode such combinatorial tasks within a QUBO framework. However, the limited qubit counts and memory capacities of current quantum hardware and simulators necessitate meticulous encoding strategies. We propose two distinct QUBO-based approaches to address these challenges. The pure quantum approach explicitly encodes all route-to-drone assignments using binary decision variables \( x_{i,\alpha} \), but this method quickly becomes infeasible for large \( m \) due to the exponential proliferation of variables. In contrast, the hybrid QUBO approach employs binary logarithmic encoding of drone indices to reduce the variable count substantially, incorporates classical post-processing to ensure solution validity, and applies a high-level load balancing constraint to maintain a manageable model size.

The hybrid scheduling process integrates quantum and classical optimization techniques. The hybrid workflow follows the steps of Algorithm \ref{alg:hybrid}. This approach balances loads while utilizing quantum optimization for initial route assignment. The classical correction guarantees that no drone remains idle and that every route is assigned to a drone.

\begin{algorithm}
\caption{Hybrid Drone Scheduling via QUBO}
\label{alg:hybrid}
\KwIn{Routes from Tabu Search, $n$ routes, $m$ drones}
\KwOut{Optimized drone assignments}

\textbf{Step 1: Formulate QUBO Model}\;
\For{each route $r_i$}{
    Encode route-destination assignments using binary variables\;
    Enforce exactly-one constraint \;
    Apply load balancing penalty\;
}

\textbf{Step 2: Solve QUBO Using QAOA}\;
Send QUBO to Quantum Approximate Optimization Algorithm (QAOA)\;
Extract initial route-to-drone assignments from quantum solution\;

\textbf{Step 3: Classical Post-Processing}\;
Identify unassigned routes from quantum output\;
\For{each unassigned route $r_i$}{
    Assign $r_i$ to the least-loaded drone\;
}

\textbf{Step 4: Ensuring All Drones Are Assigned at Least One Route}\;
Identify drones with zero assigned routes\;
\For{each unassigned drone $d_j$}{
    Find the most-loaded drone $d_k$\;
    Transfer a route from $d_k$ to $d_j$\;
}

\Return Optimized Drone Assignments\;

\end{algorithm}

After the convergence of the routing phase, we obtain a set of routes $\{r_1, r_2, \dots, r_n\}$ with associated times $\{T_1, T_2, \dots, T_n\}$. The focus then shifts to assigning these routes to $m$ drones via a QUBO formulation.

We introduce a binary variable \( x_{i,\alpha} \) for each route \( r_i \) and each drone \( d_\alpha \), defined as follows:
\begin{equation}
x_{i,\alpha} = \begin{cases}
1, & \text{if route } r_i \text{ is assigned to drone } d_\alpha, \\
0, & \text{otherwise}.
\end{cases}
\end{equation}
Each route \( r_i \) must be assigned to exactly one drone, enforcing a constraint that ensures a unique assignment across the fleet. This is expressed mathematically as:
\begin{equation}
\sum_{\alpha=1}^{m} x_{i,\alpha} = 1, \quad \forall \, i \in \{1, \dots, n\}
\end{equation}
In a QUBO formulation, this constraint is typically incorporated through a penalty term, given by:
\begin{equation}
H_{\text{assign}} = \sum_{i=1}^{n} \left( \sum_{\alpha=1}^{m} x_{i,\alpha} - 1 \right)^2
\end{equation}
Optionally, we can enforce that every drone receives at least one route to ensure full utilization of the fleet, though this condition may be relaxed for large \( m \) to maintain feasibility. This requirement is written as:
\begin{equation}
\sum_{i=1}^{n} x_{i,\alpha} \geq 1, \quad \forall \, \alpha \in \{1, \dots, m\}
\end{equation}
In QUBO form, this can be approximated with a penalty term, expressed as:
\begin{equation}
H_{\text{drone\_usage}} = \sum_{\alpha=1}^{m} \left(1 - \sum_{i=1}^{n} x_{i,\alpha}\right)_{+}^2,
\end{equation}
where \( (z)_+ = \max(z,0) \) is approximated or enforced with penalty terms in an unconstrained binary setting. To represent the makespan, we introduce a variable \( M_{\max} \) to capture the largest total load across all drones, defined by the inequality:
\begin{equation}
M_{\max} \ge \sum_{r_i \in A_\alpha} T_i, \quad \forall \, \alpha \in \{1, \dots, m\}
\end{equation}
This constraint can be approximated in the QUBO by adding penalty terms such as:
\begin{equation}
H_{\text{makespan}} = \sum_{\alpha=1}^{m} \sum_{i=1}^{n} \bigl( -C \, T_i \, x_{i,\alpha} \bigr) + \ldots
\end{equation}
ensuring \( M_{\max} \) remains sufficiently large to accommodate any assignment. The negative sign arises because QUBO solvers minimize the objective function, prompting the solution to minimize \( M_{\max} \) by maximizing negative contributions. A significant limitation of this approach is the use of \( n \times m \) binary variables \( \{x_{i,\alpha}\} \), which becomes impractical for large \( n \) and \( m \). Current quantum hardware and simulators struggle to manage such extensive binary expansions efficiently, with numerical experiments revealing exponential or near-exponential resource demands as \( m \) exceeds three or four.

To address this memory overhead, we propose a hybrid QUBO formulation that encodes the drone index \( \alpha \in \{0, 1, \dots, m-1\} \) for each route \( r_i \) using \( B = \lceil \log_2 (m) \rceil \) bits, defined as:
\begin{equation}
x_{i,b} \in \{0, 1\}, \quad 0 \le b < B
\end{equation}
The drone assignment for route \( r_i \) is then decoded via:
\begin{equation}
\text{Drone}_i = \sum_{b=0}^{B-1} 2^b \, x_{i,b}
\end{equation}
This logarithmic compression reduces the variable count from \( n \times m \) to \( n \times B \), significantly shrinking the problem dimension. We penalize invalid assignments to ensure the assigned drone index remains valid and does not exceed \( m-1 \). One approach approximates this constraint with the following:
\begin{equation}
H_{\text{exactly-one}} = \sum_{i=1}^{n} \left( 1 - \sum_{b=0}^{B-1} 2^b x_{i,b} \right)^2
\label{eq:onlyone}
\end{equation}
However, careful calibration is needed if \( \sum_b 2^b x_{i,b} \ge m \), and additional penalty terms with coefficient \( P \) can be introduced for such cases. To distribute routes evenly among drones, we define \( L_{\text{avg}} = n / m \) as the expected number of routes per drone and incorporate a balancing term:

\begin{equation}
H_{\text{balance}} = \sum_{\alpha=0}^{m-1} \Bigl( \sum_{i=1}^{n} \mathbf{1}\{\text{Drone}_i = \alpha\} - L_{\text{avg}} \Bigr)^2
\end{equation}

Here, \( \mathbf{1}\{\text{Drone}_i = \alpha\} \) indicates whether route \( i \) is assigned to drone \( \alpha \), expanded in the QUBO using the bits \( x_{i,b} \). After solving the QUBO with a quantum solver such as QAOA, the bits for each route are decoded to determine a drone index. Classical post-processing then ensures feasibility and optimizes load distribution. If some routes decode to an invalid index or multiple routes map to a non-existent drone, they are reassigned to the least-loaded drone to minimize additional cost. Additionally, if the solver yields a high load skew among drones, a local classical search or small-scale re-optimization redistributes routes for better balance. This hybrid method leverages quantum optimization to tackle core constraints while controlling variable growth, with classical corrections ensuring a feasible final assignment.

\subsection{Implementation Details}
\label{subsec:implementation}
Both QUBO formulations are mapped to the quantum annealer’s qubit topology, requiring up to one hundred qubits for problem instances of up to fifty nodes. The multi-start routing phase iterates \( k = 1000 \) times, selecting the lowest \( T \), while the scheduling phase runs once per route set. This framework balances quantum exploration with classical refinement, addressing the scalability limitations noted in Section~\ref{background}.

\section{Experiments and Results} \label{experiments}
For our experiments, we adapted the Augerat (1995) benchmark dataset \( P \), as detailed in~\cite{uchoa2017new}, to model real-world drone delivery constraints within our QUADRO framework. Initially designed for vehicle routing, this dataset was modified to reflect drone-specific logistics by scaling vehicle capacity and node demands to align with realistic drone payload limits, as specified in Table~\ref{tab:uav_parameters} (e.g., drone weight \( W = 7.5 \, \text{kg} \), payload capacity \( L = 2.5 \, \text{kg} \)). We utilized a subset of the full dataset, comprising twenty-four problems, selecting eight instances detailed in Table~\ref{tab:dataset} to evaluate performance across varying node counts and depot locations. The node demands were adjusted using Eqn. \eqref{eq:node_demand}.
\begin{equation}
r_i = 
\begin{cases} 
d_i \mod L & \text{if } d_i \mod L \neq 0 \\
1.5 & \text{otherwise}
\end{cases}
\label{eq:node_demand}
\end{equation}
where \( r_i \) denotes the demand of customer \( o_i \) in the drone routing problem, \( L \) is the drone’s payload capacity, and \( d_i \) is the original demand from dataset \( P \). This adjustment ensures demands remain feasible for drones, capping them below \( L \) while assigning a default value of 1.5 kg when the modulo operation yields zero, thereby preserving problem complexity and realism.

\begin{table}[h]
\centering
\begin{threeparttable}    
    \caption{Subset of Augerat Dataset P}
    \label{tab:dataset}
    \centering
        \begin{tabular}{lcc}
        \hline        
        \textbf{Problem} & \textbf{No. of Nodes} & \textbf{Depot Location} \\
        \hline
         P-N16-K8 & 16 & S\\
         P-N19-K2 & 19 & S\\
         P-N20-K2 & 20 & S\\
         P-N21-K2 & 21 & S\\
         P-N22-K2 & 22 & S\\         
         P-N23-K8 & 23 & S\\              
         P-N40-K5 & 40 & C\\
         P-N51-K10 & 51 & C\\
         \hline
    \end{tabular}
    \begin{tablenotes}
      \item C = Centered, S = Shifted to the left
    \end{tablenotes}    
\end{threeparttable}   
\end{table}  

\subsection{Experimental Setup}
We implemented its hybrid quantum-classical framework across a simulated quantum environment and a classical computing platform to evaluate QUADRO's performance. The quantum components, encompassing the QAOA-based local search for routing and scheduling, were simulated using Qiskit~\cite{qiskit2024}, an open-source quantum computing framework. Qiskit facilitated the construction and execution of QUBO Hamiltonians on a simulated quantum backend, mimicking NISQ-era hardware constraints with up to one hundred qubits, suitable for the problem scales tested (16–51 nodes). Experiments, including classical components and quantum simulation, were executed on an ASUS Zenbook Pro Duo equipped with an Intel Core i9-11900H processor (8 cores, 2.5 GHz base frequency) and 32 GB of RAM. This setup ran within Visual Studio Code, leveraging Python for seamless integration with Qiskit.

\subsection{Drone Delivery Experiments}

For our experiments, we evaluated QUADRO on the eight selected instances from the Augerat (1995) dataset \( P \), as detailed in Table~\ref{tab:dataset}. In the routing phase, each iteration of the QAOA-based local search involved up to \( N \times M \) quantum calls, where \( N \) is the number of nodes and \( M \) is the number of routes, corresponding to potential node-route assignments. Pre-processing filtered these assignments, excluding any that would exceed the payload capacity \( L = 2.5 \, \text{kg} \) (Table~\ref{tab:uav_parameters}), reducing computational overhead. The resulting QUBO formulations ranged from four to twenty-five binary variables, reflecting a maximum of five nodes per route (e.g., with demands of 0.5 kg each, per Eqn.~\eqref{eq:node_demand}). Post-processing ensured battery feasibility by rejecting infeasible solutions and maintaining alignment with drone constraints.

\begin{figure}[tb]
    \centering
    \includegraphics[width=0.98\linewidth]{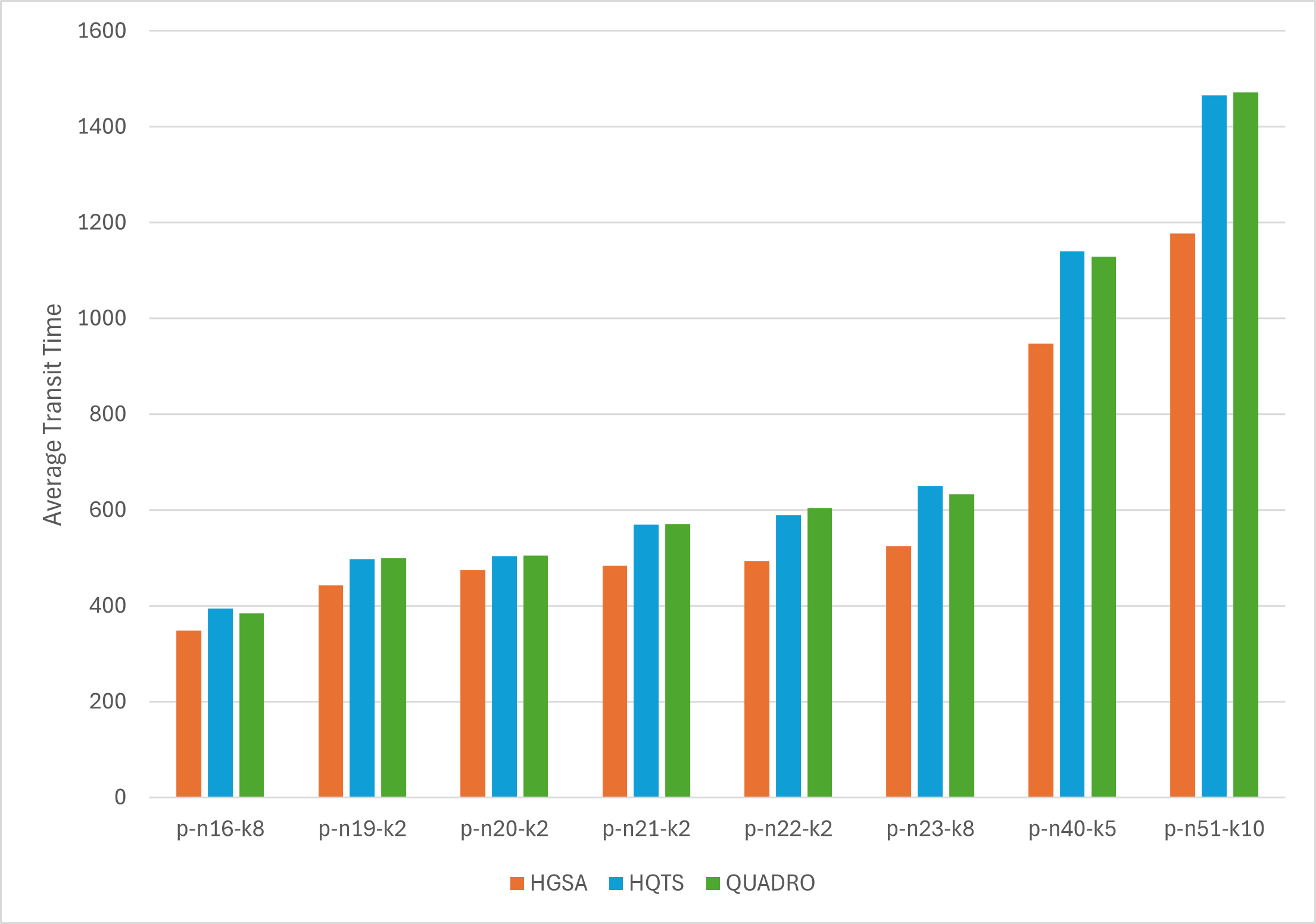}
    \caption{Comparison of average routing transit times (in minutes) across eight Augerat dataset instances for QUADRO, HQTS, and HGSA, showcasing the performance of our quantum-enhanced drone routing framework QUADRO against the prior hybrid method HQTS and a classical benchmark HGSA~\cite{sajid2022routing}. All methods exhibit rising transit times with increasing node counts, with QUADRO closely tracking HQTS while exceeding HGSA’s best-case results.}
    \label{fig:routing_comparison}
\end{figure}

Experimental runtimes within QUADRO’s routing phase scaled with the number of nodes \( N \), as larger \( N \) values drove an increase in routes \( M \), amplifying quantum calls (\( N \times M \)) and prolonging local search durations. The stochastic multi-start perturbation strategy introduced variability across runs, prompting us to execute each of the eight problems (Table~\ref{tab:dataset}) three times and report averaged results in Table~\ref{tab:routing_results}. For benchmarking, we juxtapose QUADRO against the classical Hybrid Genetic Simulated Annealing (HGSA) approach by \cite{sajid2022routing} and the prior Hybrid Quantum Tabu Search framework~\cite{holliday2024tabu}, both evaluated on the same dataset. While QUADRO’s travel times and power consumption consistently surpassed HGSA’s best-case outcomes, owing to unverified parameter alignment and inaccessible source code, it closely mirrors HQTS’s performance, reflecting shared hybrid principles. All three methods exhibit parallel trends, with travel time and power consumption escalating alongside problem complexity (e.g., node count), as evidenced in Table~\ref{tab:routing_results} and Figure~\ref{fig:routing_comparison}. This evaluation underscores QUADRO’s viability compared to a proven hybrid approach in HQTS.

\begin{table}[h]
\centering
\resizebox{\linewidth}{!}{%
\begin{threeparttable}
    \caption{Average Routing Results: QUADRO vs. HGSA and HQTS}
    \label{tab:routing_results}
    \begin{tabular}{l c c c c c c}
        \hline
        & \multicolumn{2}{c}{\textbf{HGSA}} & \multicolumn{2}{c}{\textbf{HQTS}} & \multicolumn{2}{c}{\textbf{QUADRO (Ours)}} \\
        \cline{2-3} \cline{4-5} \cline{6-7}
        \textbf{Problem} & \textbf{Time} & \textbf{Power} & \textbf{Time} & \textbf{Power} & \textbf{Time} & \textbf{Power} \\
        \hline
        P-N16-K8  & 348.66 & 1.16 & 394.53 & 1.24 & 384.76 & 1.15 \\
        P-N19-K2  & 442.96 & 1.57 & 497.72 & 1.60 & 499.70 & 1.62 \\
        P-N20-K2  & 474.45 & 1.66 & 503.60 & 1.59 & 505.03 & 1.60 \\
        P-N21-K2  & 484.05 & 1.69 & 569.55 & 1.72 & 570.40 & 1.73 \\
        P-N22-K2  & 494.23 & 1.72 & 589.68 & 1.84 & 604.65 & 1.84 \\
        P-N23-K8  & 524.91 & 1.80 & 650.21 & 1.92 & 633.22 & 1.96 \\
        P-N40-K5  & 947.07 & 3.40 & 1,139.63 & 3.52 & 1,128.88 & 3.56 \\
        P-N51-K10 & 1,176.98 & 4.21 & 1,466.08 & 4.61 & 1,471.83 & 4.67 \\
        \hline
    \end{tabular}
    \begin{tablenotes}
        \item Note: Travel times are in minutes; power consumption is in kilowatt-hours (kWh). HGSA results reflect best-case values from~\cite{sajid2022routing}; HQTS and QUADRO values are averaged over three runs.
    \end{tablenotes}
\end{threeparttable}
}
\end{table}

The drone scheduling models were evaluated for each dataset using three independent trials per drone count. The average makespan was computed from the individual makespans as formulated in Eqn. \eqref{eq:makespan}. A value of 1.25 was used for \( \tau_c \) (the time in hours to recharge the
drone) in the equation (Table~\ref{tab:uav_parameters}). Recharge time is required between consecutive route completions, assuming each drone starts fully charged, and recharge after its final assigned route is not considered.

The hybrid QUBO model successfully assigned routes and minimized makespan across all tested datasets, see Table~\ref{tab:combined}. The table shows the results for the P-N16-K8 dataset, where increasing drones resulted in a progressive decrease in makespan. A similar trend is observed in P-N19-K2, where additional drones reduced the average makespan. For datasets with higher route counts, such as P-N40-K5 and P-N51-K10, QUBO complexity limited scalability beyond four drones, preventing additional trials.

The purely quantum approach was evaluated for smaller datasets before encountering memory constraints. The results, shown in Table~\ref{tab:combined}, indicate that where solutions were feasible, the purely quantum approach consistently produced lower makespans than the hybrid method. However, due to computational limitations, the purely quantum model was unable to scale as well.

Table~\ref{tab:combined} shows that the purely quantum approach yielded lower makespan values for P-N16-K8, P-N19-K2 and P-N20-K2 compared to the hybrid method, with an observed reduction of approximately 35--45\%- for two drones. The purely quantum model was constrained to a few drones due to memory limitations, while the hybrid approach scaled across datasets with larger route counts and drone fleets.

In datasets with higher route counts, QUBO complexity growth restricted solution feasibility for both methods. The results in Table~\ref{tab:combined} illustrate the limitations, where the hybrid approach failed beyond four drones due to memory constraints, and the purely quantum approach reached memory limits beyond three drones.

\begin{table*}[t]
    \centering
    \caption{Comparison of Hybrid and Quantum Makespan Across Problem Instances}
    \label{tab:combined}
        \begin{tabular}{c | cc | cc | cc | c | c | c}
            \hline
            & \multicolumn{2}{c|}{P-N16-K8} & \multicolumn{2}{c|}{P-N19-K2} & \multicolumn{2}{c|}{P-N20-K2} & \multicolumn{1}{c|}{P-N21-K2} & \multicolumn{1}{c|}{P-N40-K5} & \multicolumn{1}{c|}{P-N51-K10} \\
            \hline
            \textbf{Drones} & \textbf{Hybrid} & \textbf{Quantum} & \textbf{Hybrid} & \textbf{Quantum} & \textbf{Hybrid} & \textbf{Quantum} & \textbf{Hybrid} & \textbf{Hybrid} & \textbf{Hybrid} \\
            \hline
            2  & 38.85 & 21.25 & 50.2  & 31.76 & 44.63 & 29.6  & 57.62 & 104.33 & 153.03 \\
            3  & 18.23 & 16.95 & 25.13 & --    & 29.22 & --    & 34.08 & 62.48  & 94.12 \\
            4  & 17.43 & --    & 28.87 & --    & 33.18 & --    & 27.43 & 73.88  & 86.62  \\
            5  & 14.20 & --    & 18.35 & --    & 18.40 & --    & 26.87 & --     & --     \\
            6  & 10.03 & --    & 17.70 & --    & 15.80 & --    & 18.72 & --     & --     \\
            7  & 8.20  & --    & 12.72 & --    & 14.52 & --    & 17.58 & --     & --     \\
            8  & --    & --    & 11.63 & --    & 12.63 & --    & 16.50 & --     & --     \\
            9  & --    & --    & 8.60  & --    & 8.80  & --    & 19.62 & --     & --     \\
            10 & --    & --    & --    & --    & --    & --    & 10.08 & --     & --     \\
            11 & --    & --    & --    & --    & --    & --    & 8.60  & --     & --     \\
            \hline
        \end{tabular}
\end{table*}

\section{Conclusions}\label{conclusion}

This work has demonstrated the efficacy of QUADRO, a hybrid quantum-classical framework, in addressing the dual challenges of drone routing and scheduling within the constraints of near-term quantum computing. For the EC-CUAVRP, QUADRO employs a multi-start heuristic with a QAOA-based local search, achieving competitive transit times compared to classical benchmarks like HGSA and our prior HQTS approach, as shown in Table~\ref{tab:routing_results}. By formulating routing as a QUBO problem, QUADRO optimizes route configurations under payload and battery limits, leveraging quantum exploration to enhance solution diversity while relying on classical validation to ensure feasibility. This synergy yields transit times that closely track HQTS but fall short of the purely classical approach of HGSA, highlighting the potential of quantum-assisted routing for real-world drone logistics.

In the UAVSP, QUADRO’s hybrid QUBO approach significantly advances scalability over a purely quantum model. The pure quantum formulation, encoding assignments with \( n \times m \) variables, delivers lower makespans for small fleets (e.g., 21.25 vs. 38.85 hours for P-N16-K8 with two drones) but falters beyond three drones due to exponential resource demands. Conversely, the hybrid method compresses variables to \( n \times \log_2(m) \), enabling scheduling for larger fleets (up to eleven drones in P-N21-K2) with moderate overhead, as shown in Table~\ref{tab:combined}. Classical post-processing corrects invalid assignments and balances loads, ensuring practical feasibility. This scalability, coupled with makespan reductions as drone counts increase, underscores QUADRO’s adaptability to complex scheduling scenarios.

QUADRO’s achievements reflect a balanced integration of quantum optimization and classical refinement, bridging the gap between NISQ-era limitations and operational drone fleet management. The framework’s routing success lies in its ability to explore diverse, time-efficient routes, while its scheduling prowess stems from logarithmic encoding and post-processing, mitigating quantum hardware constraints. Compared to other methods, QUADRO offers a compelling alternative, achieving comparable or superior performance with fewer than one hundred qubits. These results affirm the promise of hybrid quantum-classical methods for combinatorial optimization in logistics.

Future work will extend QUADRO’s capabilities in several directions. Enhancing the QAOA-based local search with deeper circuits or alternative quantum algorithms, such as variational quantum eigensolvers, may improve routing precision for larger node sets. Incorporating real-time environmental factors, e.g., wind or environment conditions, into the problem formulations could further align QUADRO with practical deployment. Exploring quantum annealing alongside QAOA might yield more robust assignments for scheduling, particularly for heterogeneous fleets with varying drone capacities. In addition, integrating machine learning to predict perturbation strategies in the multi-start routing phase could accelerate convergence, reducing computational overhead. Finally, validating QUADRO on physical quantum hardware, rather than simulators, would provide critical insights into its performance under NISQ conditions, advancing its transition from theoretical framework to operational tool.

{
    \small
    \bibliographystyle{IEEEtran}
    \bibliography{IEEE_quantum_drones}

\begin{thebibliography}{10}
\providecommand{\url}[1]{#1}
\csname url@samestyle\endcsname
\providecommand{\newblock}{\relax}
\providecommand{\bibinfo}[2]{#2}
\providecommand{\BIBentrySTDinterwordspacing}{\spaceskip=0pt\relax}
\providecommand{\BIBentryALTinterwordstretchfactor}{4}
\providecommand{\BIBentryALTinterwordspacing}{\spaceskip=\fontdimen2\font plus
\BIBentryALTinterwordstretchfactor\fontdimen3\font minus \fontdimen4\font\relax}
\providecommand{\BIBforeignlanguage}[2]{{%
\expandafter\ifx\csname l@#1\endcsname\relax
\typeout{** WARNING: IEEEtran.bst: No hyphenation pattern has been}%
\typeout{** loaded for the language `#1'. Using the pattern for}%
\typeout{** the default language instead.}%
\else
\language=\csname l@#1\endcsname
\fi
#2}}
\providecommand{\BIBdecl}{\relax}
\BIBdecl

\bibitem{sajid2022routing}
M.~Sajid, H.~Mittal, S.~Pare, and M.~Prasad, ``Routing and scheduling optimization for uav assisted delivery system: A hybrid approach,'' \emph{Applied Soft Computing}, vol. 126, p. 109225, 2022.

\bibitem{gao2021weather}
M.~Gao, C.~H. Hugenholtz, T.~A. Fox, M.~Kucharczyk, T.~E. Barchyn, and P.~R. Nesbit, ``Weather constraints on global drone flyability,'' \emph{Scientific reports}, vol.~11, no.~1, p. 12092, 2021.

\bibitem{tran2022management}
T.-H. Tran and D.-D. Nguyen, ``Management and regulation of drone operation in urban environment: A case study,'' \emph{Social Sciences}, vol.~11, no.~10, p. 474, 2022.

\bibitem{boysen2018drone}
N.~Boysen, D.~Briskorn, S.~Fedtke, and S.~Schwerdfeger, ``Drone delivery from trucks: Drone scheduling for given truck routes,'' \emph{Networks}, vol.~72, no.~4, pp. 506--527, 2018.

\bibitem{kim2020drone}
J.~Kim, H.~Moon, and H.~Jung, ``Drone-based parcel delivery using the rooftops of city buildings: Model and solution,'' \emph{Applied Sciences}, vol.~10, no.~12, p. 4362, 2020.

\bibitem{torabbeigi2020drone}
M.~Torabbeigi, G.~J. Lim, and S.~J. Kim, ``Drone delivery scheduling optimization considering payload-induced battery consumption rates,'' \emph{Journal of Intelligent \& Robotic Systems}, vol.~97, pp. 471--487, 2020.

\bibitem{shivgan2020energy}
R.~Shivgan and Z.~Dong, ``Energy-efficient drone coverage path planning using genetic algorithm,'' in \emph{2020 IEEE 21st International Conference on High Performance Switching and Routing (HPSR)}.\hskip 1em plus 0.5em minus 0.4em\relax IEEE, 2020, pp. 1--6.

\bibitem{hoffman2013traveling}
K.~L. Hoffman, M.~Padberg, G.~Rinaldi \emph{et~al.}, ``Traveling salesman problem,'' \emph{Encyclopedia of operations research and management science}, vol.~1, pp. 1573--1578, 2013.

\bibitem{toth2002vehicle}
P.~Toth and D.~Vigo, \emph{The vehicle routing problem}.\hskip 1em plus 0.5em minus 0.4em\relax SIAM, 2002.

\bibitem{preskill2018quantum}
J.~Preskill, ``Quantum computing in the nisq era and beyond,'' \emph{Quantum}, vol.~2, p.~79, 2018.

\bibitem{feld2019hybrid}
S.~Feld, C.~Roch, T.~Gabor, C.~Seidel, F.~Neukart, I.~Galter, W.~Mauerer, and C.~Linnhoff-Popien, ``A hybrid solution method for the capacitated vehicle routing problem using a quantum annealer,'' \emph{Frontiers in ICT}, vol.~6, p.~13, 2019.

\bibitem{borowski2020new}
M.~Borowski, P.~Gora, K.~Karnas, M.~B{\l}ajda, K.~Kr{\'o}l, A.~Matyjasek, D.~Burczyk, M.~Szewczyk, and M.~Kutwin, ``New hybrid quantum annealing algorithms for solving vehicle routing problem,'' in \emph{International Conference on Computational Science}.\hskip 1em plus 0.5em minus 0.4em\relax Springer, 2020, pp. 546--561.

\bibitem{sinno2023performance}
S.~Sinno, T.~Gro{\ss}, A.~Mott, A.~Sahoo, D.~Honnalli, S.~Thuravakkath, and B.~Bhalgamiya, ``Performance of commercial quantum annealing solvers for the capacitated vehicle routing problem,'' \emph{arXiv preprint arXiv:2309.05564}, 2023.

\bibitem{holliday2024tabu}
J.~B. Holliday, B.~Morgan, H.~Churchill, and K.~Luu, ``Hybrid quantum tabu search for solving the vehicle routing problem,'' in \emph{2024 IEEE international conference on quantum computing and engineering (QCE)}.\hskip 1em plus 0.5em minus 0.4em\relax IEEE, 2024, pp. 353--358.

\bibitem{XBacWK2}
X.-B. Nguyen, S.~Khan, H.~Churchill, and K.~Luu, ``Quantum vision clustering,'' in \emph{arXiv:2309.09907}, 2025.

\bibitem{XBacWK1}
X.-B. Nguyen, H.-Q. Nguyen, S.~Y.-C. Chen, S.~Khan, H.~Churchill, and K.~Luu, ``Qclusformer: A quantum transformer-based framework for unsupervised visual clustering,'' in \emph{IEEE International Conference on Quantum Computing and Engineering}.\hskip 1em plus 0.5em minus 0.4em\relax IEEE, 2024.

\bibitem{XBacjournal1}
X.-B. Nguyen, H.-Q. Nguyen, H.~Churchill, S.~Khan, and K.~Luu, ``Quantum visual feature encoding revisited,'' \emph{Quantum Machine Intelligence Journal}, vol.~6, no.~61, 2024.

\bibitem{AdityaQIP}
A.~Dendukuri and K.~Luu, ``Image processing in quantum computers,'' in \emph{arXiv:1812.11042}, 2018.

\bibitem{kumar2022futuristic}
A.~Kumar, D.~A. de~Jesus~Pacheco, K.~Kaushik, and J.~J. Rodrigues, ``Futuristic view of the internet of quantum drones: review, challenges and research agenda,'' \emph{Vehicular Communications}, vol.~36, p. 100487, 2022.

\bibitem{farhi2014quantum}
E.~Farhi, J.~Goldstone, and S.~Gutmann, ``A quantum approximate optimization algorithm,'' \emph{arXiv preprint arXiv:1411.4028}, 2014.

\bibitem{wang2018quantum}
Z.~Wang, S.~Hadfield, Z.~Jiang, and E.~G. Rieffel, ``Quantum approximate optimization algorithm for maxcut: A fermionic view,'' \emph{Physical Review A}, vol.~97, no.~2, p. 022304, 2018.

\bibitem{qian2023comparative}
W.~Qian, R.~A. Basili, M.~M. Eshaghian-Wilner, A.~Khokhar, G.~Luecke, and J.~P. Vary, ``Comparative study of variations in quantum approximate optimization algorithms for the traveling salesman problem,'' \emph{Entropy}, vol.~25, no.~8, p. 1238, 2023.

\bibitem{ramezani2024reducing}
M.~Ramezani, S.~Salami, M.~Shokhmkar, M.~Moradi, and A.~Bahrampour, ``Reducing the number of qubits from $n^{2}$ to $n \log_2(n)$ to solve the traveling salesman problem with quantum computers: A proposal for demonstrating quantum supremacy in the nisq era,'' \emph{arXiv preprint arXiv:2402.18530}, 2024.

\bibitem{choi2020quantum}
J.~Choi, S.~Oh, and J.~Kim, ``Quantum approximation for wireless scheduling,'' \emph{Applied Sciences}, vol.~10, no.~20, p. 7116, 2020.

\bibitem{kurowski2023application}
K.~Kurowski, T.~Pecyna, M.~Slysz, R.~R{\'o}{\.z}ycki, G.~Walig{\'o}ra, and J.~W\c{e}glarz, ``Application of quantum approximate optimization algorithm to job shop scheduling problem,'' \emph{European Journal of Operational Research}, vol. 310, no.~2, pp. 518--528, 2023.

\bibitem{vista2023hybrid}
F.~Vista, G.~Iacovelli, and L.~A. Grieco, ``Hybrid quantum-classical scheduling optimization in uav-enabled iot networks,'' \emph{Quantum Information Processing}, vol.~22, no.~1, p.~47, 2023.

\bibitem{hua2024quantum}
R.~Hua, D.~Di~Lorenzo, F.~Chinesta, and P.~Codognet, ``Quantum annealing solutions for drone route planning problems,'' in \emph{2024 IEEE International Conference on Quantum Computing and Engineering (QCE)}, vol.~1.\hskip 1em plus 0.5em minus 0.4em\relax IEEE, 2024, pp. 600--610.

\bibitem{osaba2025solving}
E.~Osaba, P.~Miranda-Rodriguez, A.~Oikonomakis, M.~Petri{\v{c}}, A.~Ruiz, S.~Bock, and M.-A. Kourtis, ``Solving drone routing problems with quantum computing: A hybrid approach combining quantum annealing and gate-based paradigms,'' \emph{arXiv preprint arXiv:2501.18432}, 2025.

\bibitem{clarke1964scheduling}
G.~Clarke and J.~W. Wright, ``Scheduling of vehicles from a central depot to a number of delivery points,'' \emph{Operations research}, vol.~12, no.~4, pp. 568--581, 1964.

\bibitem{lucas2014ising}
A.~Lucas, ``Ising formulations of many np problems,'' \emph{Frontiers in physics}, vol.~2, p. 74887, 2014.

\bibitem{uchoa2017new}
E.~Uchoa, D.~Pecin, A.~Pessoa, M.~Poggi, T.~Vidal, and A.~Subramanian, ``New benchmark instances for the capacitated vehicle routing problem,'' \emph{European Journal of Operational Research}, vol. 257, no.~3, pp. 845--858, 2017.

\bibitem{qiskit2024}
A.~Javadi-Abhari, M.~Treinish, K.~Krsulich, C.~J. Wood, J.~Lishman, J.~Gacon, S.~Martiel, P.~D. Nation, L.~S. Bishop, A.~W. Cross, B.~R. Johnson, and J.~M. Gambetta, ``Quantum computing with {Q}iskit,'' 2024.

\end{thebibliography}
}

\end{document}